# EEG Based Emotion Sensing using deep learning convolutional neural networks

First A. Shivaditya Shivganesh

*Abstract*— Deep Learning has impacted various fields especially in bio-medical applications. The deep learning algorithms work well in both structured and unstructured data. Especially, convolutional neural network work well signal based data like EEG data. These types of data may or may not follow a pattern in their data. Algorithms like CNN help feature engineering and simplistic interpretation of the data. These algorithms also better in comparison to other algorithms when generalized to a data belonging to larger data set.

## I. INTRODUCTION

THE concept of emotion sensing using EEG or Electro Encephalography is a novel idea often applied in BCI or Brain Computer Interfaces applications. Emotion sensing or classification often involves usage of deep learning algorithms like Autoencoders and Sparse networks but the approach can be simplified for lesser usage of computational resources.

Brain Computer Interface is one the most advanced field in Computational Neuroscience. It provides a pathway for brain to interact with computer. BCI technologies have found their application in various fields like Motor Signal classification, disease prediction using EEG data. EEG data is extremely useful in diagnosing or predicting diseases like Schizophrenia and predict conditions like seizures.

There are two forms of signals based on which emotions are classified in BCI application one is non-physiological and other is physiological. The non-physiological one is based on facial expressions. Similarly, the physiological signals are grouped into two types electrical (Electro Corticogram (ECog), Electro Encephalogram (EEG), Magneto Encephalogram (MEG)) and hemodynamics (functional Near-Infrared Spectroscopy (fNIRS), functional Magnetic Resonance Imaging (fMRI)) [19].

EEG being non stationery in nature it is most appropriate to use time-frequency domain methods [17].

There are several ways to extract information from EEG data like autoregressive (AR) methods, Phase reconstruction method and others [19].

We have tried to classify the emotional responses produced by the individual in study with response to auditory stimuli into multiple classes using deep learning algorithms.

Less information is available about the characteristics of positive emotions as it is significantly difficult to pin-point and classify the positive emotions.[11]

Often analysis or classification studies on emotions use Arousal- valence model for classification this hides a lot of information about each individual positive emotion. As it is much easier to classify the emotions as Arousal than individual positive emotions [11]. Rise of studies in positive psychology has inspired more exploration in positive emotions [7]. It is believed that positive emotions broaden people's cognitive skills and improve the attention scope of the person [7]. Even though the positive emotions play an important part in the well-being of the person not much is known about its physiological mechanisms. It so because they elicit less reactivity that is, they do not show much cardiovascular response and they do not differ much from the baseline cardiovascular response [11].

Finding the portion of the brain responsible for the emotions is much more difficult to find in positive emotions than in negative emotions. Till date only for one positive emotion the location or source of the emotion has been identified it is the emotion "Happiness". It was located to Rostral-anterior cingulate cortex [11].

Emotions are often a complex pattern of arousal and subjective feeling and cognitive interpretation. Emotion is highly subjective field as the degree of emotion felt and quality of emotion felt vary from person to person [26]. Automatic nervous system plays an important role in emotion generation as well as regulation. Somatic nervous system also plays an important role in emotion process.

It is believed that neuro-physiological activation takes place in thalamus, limbic system and cerebral cortex. Also, extensive injury to these portions cause an impairment in emotions [27].

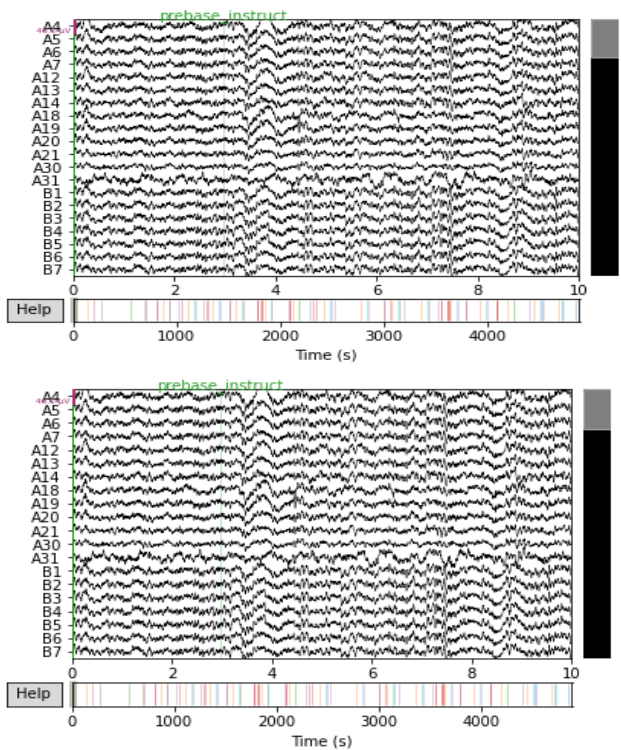

Fig. 1. Magnetization as a function of applied field. Note that "Fig." is abbreviated. There is a period after the figure number, followed by two spaces. It is good practice to explain the significance of the figure in the caption.

Earlier, physiological theory of emotions was based on the fact that environment stimuli, elicit physiological responses from viscera which is then associated with muscle movements in the body. The inference made by this theory is that for a particular stimulus there exists a particular physiological change. This theory is also called as James-Lange Theory of emotion. Another theory proposed by Cannon and Bard which claims the entire process of emotion is mediated and regulated by thalamus portion of the brain. After perception of an emotion provoking stimuli, the thalamus conveys the information to the cerebral cortex. The cerebral cortex then determines the nature of stimuli by comparing it with past experiences [26].

Automatic nervous systems or ANS play a very important role in emotion generation and regulation. It is divided into sympathetic and para-sympathetic. The ANS controls the physiological factors on the body like heart rate, blood pressure, body temperature. Sympathetic ANS prepares the body for fighting response that is, it induces a physiological arousal state ideal for a fight response [26]. On the other hand, Para-Sympathetic ANS calms the person and brings back the individual to resting state. The agnostic nature of the ANS is essential in controlling and regulating emotional responses of the body.

Hypothalamus plays a vital role in emotion regulation. It regulates the homeostatic balance. It also controls the activity and secretion of endocrine glands.[26] To some extent these glands generate hormones which are also essential in emotion regulation. Acetylcholine is essential for proper function of the neuro-transmitters. It controls the ability to create memories or remember things. Similarly, the hormone Norepinephrine is produced by adrenal gland. This hormone produces a calming emotion. This hormone is also called as the stress hormone. The hormone Dopamine is critical in producing the emotion of happiness. The hormone Endorphin is produced by pituitary gland. it reduces the feeling of pain and reduces anxiety. The hormone Serotonin is responsible of feeling of anger and urge for violence.[25] The hormone Oxytocin is responsible for emotion of affection and empathy. Along with the thalamus, the limbic system plays a important role In regulation of emotion. Amygdala is a part of limbic system. It is responsible for forming emotional memories. Cortex portion of the brain is deeply involved with emotions. The right frontal hemisphere is associated with negative emotion on the other hand the left frontal portion is involved with the positive emotions.[26]

*Stress and Emotions*

It is believed that certain emotions like anger, shame and anxiety are caused due to harmful and threating conditions. According to Lazarus's theory of stress, the stress experienced by a person is when demands exceeded the personal and social resources the individual can mobilize. This is called transactional model of stress and coping. This shows that negative emotions have high correlation with stress [27].

## II. OVERVIEW OF DEEP LEARNING NETWORKS

Deep learning often involves usage of neural networks with one or more hidden layers. The number of the hidden layers increase the depth of the neural network. One of the most predominant type of deep neural network is Convolutional Neural Networks. They have input feature matrix with which the convolutional function is performed with the with an array called Filter or Kernel. These operation help capturing spatial and temporal dependencies We can use additional parameters like step size for altering the final matrix obtained after performing the operation. Often Deep neural networks have a Max Pooling layers. These reduce the size of the image; these are helpful for reducing the usage of the computational resources. It also increases the stability of the network.

## III. METHODOLOGY

The dataset used for this research has 34 individuals. For each individual EEG data is measured and events of emotion are recorded. Using the EEG data for each individual wavelet are formed. For a single person's data multiple wavelets of 20 seconds were cut and these were used for Multi-Class classification using convolutional neural network. Using the trained model, I tried to predict class of an EEG wavelet.

*About the Dataset Used*

In this paper, I have used a publicly available dataset. This dataset was collected by Julie et al. [18]. In this dataset tasks were given to each subject participating in the study. In each task imagination of emotional states were encouraged and using a set of pre-recorded verbal suggestion emotions were suggested to the subject. Each task began with a period of relaxation [18]. It lasted for 2 mins between each task. The task began with an explanation of the emotion which was followed by five minutes of verbal guidance

A series of 15 prerecorded emotion imagery scenarios were used. Each of the recording depicted a particular emotional scenario [18]. In each scenario the subject was asked to take as much time they needed to recall or imagine that particular scenario.

A total of 250 scalp-attached electrodes were used for EEG data recording, 4 intra-ocular and 2 ECG electrodes were used, along with the EEG electrodes. A A/D resolution of 24bit was taken. The positions of each electrode on the skull of the subject were noted. The eye blink or eye movements or muscle movements artifacts were not removed from the data [18].

*Process followed*

From the dataset individual EEG data was taken. In the dataset the information about the event is mentioned. Using it I created multiple wavelets that were centered around the event and had a width of 20 seconds. No other changes were made and the data was just cut into multiple wavelets. On each wavelet Hann function was applied. Now on the transformed data Fast Fourier Transform was applied. The range of theta frequency in EEG is 4Hz to 8hz for alpha it is 8Hz to 12Hz and for gamma it is 12 Hz to 40 Hz. I grouped FFT based on the frequency and took the mean in each group. Using the locations of the electrodes used for recording the EEG data we represented them on a 2D plane by eliminating the Z Axis data. Using that I interpolated the data for more data points in each time frame. I used cubic polynomial or spline interpolation to avoid Runge's phenomenon. As it would cause oscillation at the edges of that time interval. It occurs because the EEG data is equispaced in a particular time frame. So, we use cubic interpolation instead of linear interpolation as linear interpolation would cause discontinuity in the data. Whereas, the Cubic interpolated data offers true continuity between the segment of interpolated data. This data was interpreted in form of 30x30 image with multiple colors and 3 channels.

*Hann window*

A Hann Function with length of 'a' is used to perform Hann Smoothing on the given data.

In signal processing, a function is called as window function if it is zero valued outside the selected interval. The window function is also called as apodization function or tapering function. Usually it is symmetric around the middle portion and is maximum near the middle too.

When the data directly captured from the sensors is used for preforming DFT, it causes spectral leakage. This makes it difficult to find the actual frequency of the signal. So, to avoid spectral leakage we use window function. Also, the data is non-periodic in nature as we have raw EEG data over the entire time interval but the data is split into 20 seconds frames centered around the event. So, the usage of the window becomes even more essential.

Since the noise from muscle and eye movement that is, electrical signals from facial muscles or maxillofacial muscles like orbicularis *frontalis* or the zygomaticus which contribute to noise during EEG measurement[28].

Hann function often have impact on the amplitude accuracy and frequency resolution but in comparison to other window function they have moderately low impact on the amplitude accuracy and frequency resolution.

The Hanning window starts at a value of zero and ends at a value of zero, and at its peak it has a value of one. The gradual change between 0 and 1 ensures a smooth change in amplitude.

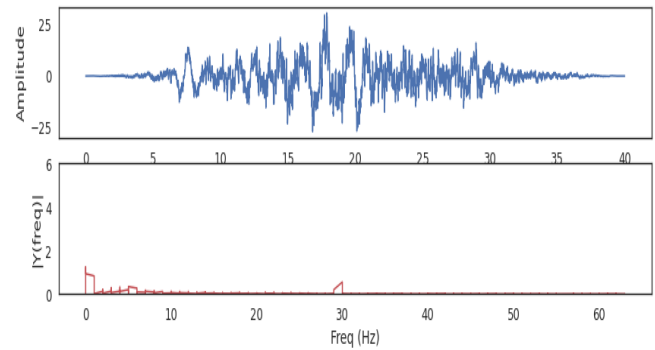

*Discrete Fourier Transform*

Discrete Fourier Transform or DFT converts the signal from its original domain to frequency domain. We get DFT by decomposing the signal sequence into individual components of different frequencies.

DFT based measurements are susceptible to spectral leakage. This occurs when DFT is computed over a block of data which is non periodic in nature.

To eliminate we choose Hann function or Hanning window. If windowing function is not correctly applied there may be errors which may be introduced into DFT. Windowing function minimizes the effect of spectral leakage. When windowing function is applied it gives rise to window's characteristics know as side lobe. Lower side lobe reduces the leakage in measured DFT but increases the bandwidth of the major lobe. Since hann function provides optimal frequency resolution and amplitude accuracy but it reduces the spectral leakage to the best extent. Welch's windowing function could be also used. Overlaps in the windows are neglected . Since it's effect on measurement time is not much. For performing DFT on a large dataset we use the FFT. But since the ordinary libraries are slow for a dataset with large datapoints. We make use of CUDA acceleration for faster calculations of DFT. We convert the two sided power spectrum to single sided power spectrum because most of the data In the real world that is most of the real world frequency data is only concerned along the positive half of the frequency spectrum and the negative frequency information becomes redundant.

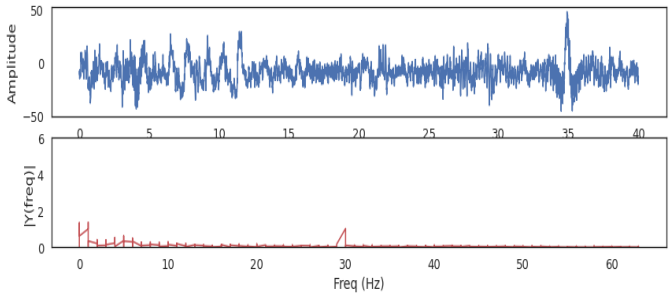

## IV. RESULTS

Using the dataset , the image pipeline was applied and for 34 subject a total of 50135 images were obtained. Each image sis of size 32px by 32px with 3 channels.

*Image generation*

The image generation process was a resource intensive one it took more than 12 hours on a dual core CPU. But when CUDA accelerated DFT calculations were performed on NVIDA K80 the entire process was completed in less than 6 hours.

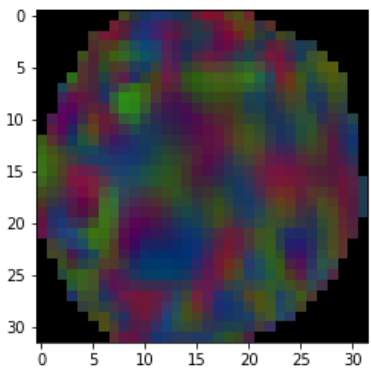

The labels were categorically encoded. High-level API like TensorFlow was used for designing the neural network. I have used a relatively small network for classification

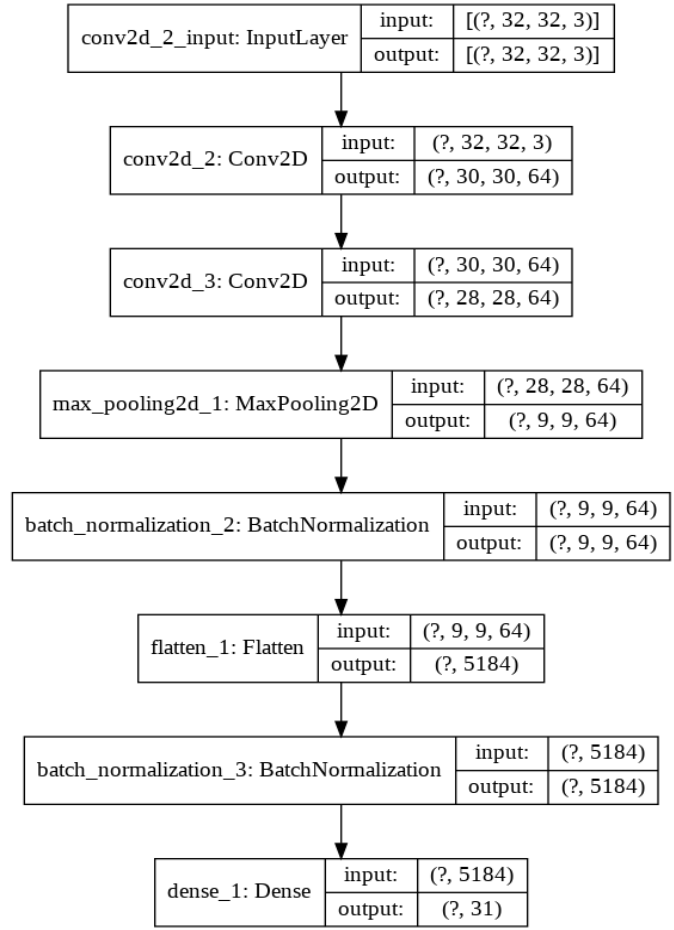

*Neural Network*

I have used TensorFlow to create the neural network. It has input size of image as (32,32,3) with kernel or filter size of 3x3. I have built a sequential model with final dense layer have Softmax activation and RMSprop as optimizer

*Training*

The image data so obtained was split into training, testing, validation dataset. The model was trained on TPU's for faster training time. It took 3 Hours to train on a Google Colab TPU completely. It obtained a training accuracy score of about 90.237% and validation accuracy of 71%

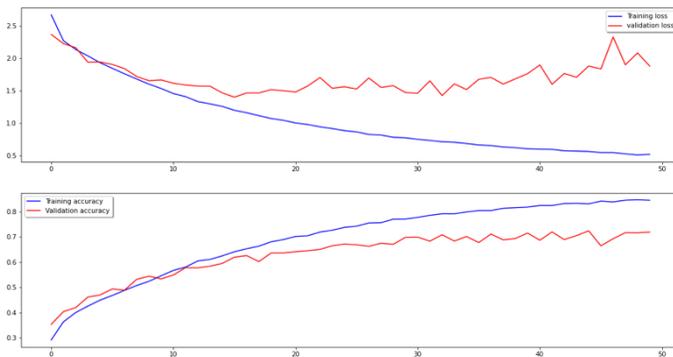

*Validation*

On a completely un seen dataset the model obtained a accuracy of about 72.03%. The validation set contained several examples from each class.

## V. CONCLUSION

The proposed EEG Classification method will be helpful in BCI application. It can be concluded that using this approach deep learning models can predict feelings in humans to a reasonable accuracy. The results are encouraging. This accuracy can we improved further by using advanced deep learning techniques which will only improve the accuracy further and enable a higher level of performance in emotion sensing.

a